\documentstyle[aps,prl,twocolumn,epsf]{revtex}
\draft
\begin{document}
\title{
Magnetic field induced charge instabilities in weakly coupled superlattices}
\author{R. Aguado$^{a}$ and G. Platero}
\address{
Departamento de Teor\'{\i}a de la Materia Condensada, Instituto de
Ciencia de Materiales de Madrid, CSIC, Cantoblanco 28049, Madrid,
Spain.}
\maketitle
\thispagestyle{empty}
\begin{abstract}
\widetext
Using a time dependent selfconsistent model
 for vertical sequential tunneling, we 
study the appearance of charge
 instabilities that lead to the formation
of electric field domains
in a weakly coupled doped superlattice in the presence of high
magnetic fields parallel to the transport direction.
The interplay between the high non linearity of the system --coming from
the Coulomb interaction-- and
the inter-Landau-level scattering at the domain walls 
(regions of charge accumulation inside the superlattice) gives
rise to new unstable negative differential conductance regions and extra
stable branches in the sawtooth-like I-V curves.
\end{abstract}
\narrowtext
Weakly coupled doped semiconductor
 superlattices are an example of non linear systems
in which all intrinsic properties related to high non linearity such
as  
multistability, spatio-temporal chaos, etc, can be externally modified
 : e.g. by the application of a bias voltage
or by the variation of the doping
densities in the wells or in the contacts.
The strong non linear transport that results from the
Coulomb interaction in
these systems presents a rich variety
of new physical phenomena:  
multistability and electric field domains formation \cite{Gra,Kwo,Wac}, 
self sustained oscillations \cite{Kas},
bifurcation to chaos \cite{Bul}, etc. 
In particular, the current flowing
through a weakly coupled doped semiconductor superlattice
 presents a complicated sawtooth
structure with unstable regions of negative differential conductance and
multistable regions. This structure of branches comes from the
 charge instabilities that appear due to the motion of the domain
 wall (acummulation layer) from one well 
to another. This motion, which is due to resonant 
tunneling between adjacent 
subbands,
leads to the formation of
electric field domains.   
In this Letter we study theoretically
 for the first time
 this phenomenon in the presence of high magnetic fields 
parallel to the
transport direction. New physical questions that were not relevant 
in the absence of
magnetic fields can be raised. In particular, what happens to the charge 
instabilities in the presence of the new energy scale in the problem, 
$\hbar\omega_{c}$?\\
Resonant tunneling through double barriers
\cite{Lea,Boe,Fer,Leo,Zou} and superlattices \cite{Mul} in the presence of
high magnetic fields has been widely studied.
Also, dynamical instabilities and bifurcations to
chaos induced by a magnetic field  
in a double barrier resonant tunneling diode 
have been recently predicted \cite{Ore}. 
It is well known that the application of a magnetic field
 perpendicular to a two dimensional electron gas
 produces the formation of Landau levels. 
 For ideal samples, the
tunneling through the heterostructure conserves the Landau level index.
However, \pagebreak\vspace*{1.6in} due to interface roughness, impurity 
scattering, or phonon
scattering, these conservation rules are relaxed and inter Landau level
transitions take place \cite{Lea,Boe,Fer,Leo,Zou}.
Recent magnetotransport experiments on weakly coupled doped
semiconductor superlattices performed
by Schmidt {\it et al}.\cite{Smi} show
new unstable regions of negative differential conductance and 
extra stable branches in the I--V curves
for a certain range of magnetic fields. In this work we will show how the
interplay between the high non linearity of the system, and 
inter-Landau-level transitions through regions of charge accumulation
in the structure --domain walls-- can explain these new instabilities.\\
The main ingredients of our model are the following:
 we assume that the characteristic time for
intersubband relaxation due to scattering
which for optical phonon scattering is about 0.1 ps, 
is much smaller than the
tunneling time. The latter is less than 0.5 ns, which is in turn much
smaller than the dielectric relaxation time responsible for
reaching a steady state, which is about 10 ns for the 
superlattices discussed in Ref.~\onlinecite{Kas}. This separation
of time scales, as well
as the configuration of a typical
sample, allows us to assume that only the ground
state
of each well is populated and that the
tunneling processes are stationary. These
assumptions then justify the use of rate equations for the
electron densities at each well.
 In order to include
relaxation in the transport direction
due to the different scattering mechanisms, we suppose
that the spectral densities in the quantum wells are Lorentzians whose
half width $\gamma$
is a parameter related to the scattering life time,
which is of the order of picoseconds.
Relaxation in the planes perpendicular to the transport
is considered imposing current conservation in the
stationary limit.
This condition determines the sequential
 current as well as the Fermi energies
within each well $\{\epsilon_{\omega_{i}}\}$ \cite{Jon,Agu}.
The electrons are described by a Fermi-Dirac distribution
in each well, which implicitly assumes that some scattering mechanism
thermalizes them towards a local equilibrium situation.
 In the absence of such a mechanism for thermalizing
the electrons the distribution functions are nonequilibrium objects 
and the rate equation is equivalent to a 
quantum kinetic method --i.e, Keldysh or Kadanoff-Baym, for
example. 
The rate equations --for a given set 
$\{\epsilon_{\omega_{i}}\}$-- are:
\begin{equation}
\frac{dn_{i}}{dt} = J_{i-1,i}-J_{i,i+1}
\hspace{2cm} i=1,\ldots,N.\label{rate}
\end{equation}
In this equation, $N$ is the number of wells, $J_{i,i+1}$
are the interwell currents and $J_{0,1}$ and $J_{N,N+1}$ are the currents
through the contacts (emitter and collector regions). 
These currents are
calculated by means of the Tunneling Hamiltonian method \cite{Jon}, and
have dependencies $J_{i,i+1} \equiv J_{i,i+1}(\epsilon_{\omega_{i}},
\epsilon_{\omega_{i+1}},\{\Phi\})$.
$\{\Phi\}$ denotes the set of
variables coming from the electrostatics: potential drops at the
accumulation and depletion layers, barriers
and wells, etc. These must be calculated
selfconsistently for each applied bias.
More details of the selfconsistent procedure
and of the electrostatic model considered here are given
in  Ref.~\onlinecite{Agu}.
These rate equations, which include all of the relevant dynamics
of the selfconsistent problem, can be rewriten in terms of 
an Amp\`ere's law that explicitly shows the time-dependent
current consisting of a displacement current plus a tunneling term
\cite{Agu}
${\varepsilon\over d}\frac{dV_{i}}{dt} + J_{i-1,i} = J(t)$.
Here $\frac{V_{i}}{d}$ is the electric field at the $i$-th barrier and
$\varepsilon$ is the GaAs static permittivity. 
Nonetheless, in this work we restrict ourselves to the
stationary regime $\frac{dn_{i}}{dt} \rightarrow 0$.
In this limit, 
the interwell currents and
the currents from the contacts are all
equal to the total sequential current $J$.
Following Ref.~\onlinecite{Agu} and considering the formation of Landau levels
in the planes perpendicular to the transport, we
obtain the following expressions for the tunneling currents: 
\begin{eqnarray}
J_{0,1}&=&\frac{e2g\hbar}{m^{*}2\pi}\sum_{j=1}^{m}
\sum_{n=0}^{n^{1}_{max}}
\int A_{Cj}^{1}(\epsilon_{z})B_{1,2}(\epsilon_{z}) 
T_{1}(\epsilon_{z})\nonumber\\
&\times&\left[f_{\epsilon_{F}}(\epsilon_{z})-f_{\epsilon_{\omega_{1}}}
(\epsilon_{z})\right]
d\epsilon_{z},\nonumber\\
J_{i,i+1}&=&\frac{e2g\hbar^{3}}{2\pi {m^{*}}^{2}}\sum_{j=1}^{m}
\sum_{n=0}^{n^{i}_{max}}
\int A_{C1}^{i}(\epsilon_{z})A_{Cj}^{i+1}(\epsilon_{z})\nonumber\\ 
&\times&B_{i,i+1}(\epsilon_{z})B_{i+1,i+2}(\epsilon_{z})
T_{i+1}(\epsilon_{z})\nonumber\\
&\times&\left[f_{\epsilon_{\omega_{i}}}(\epsilon_{z})
-f_{\epsilon_{\omega_{i+1}}}(\epsilon_{z})\right]
d\epsilon_{z},\nonumber\\
J_{N,N+1}&=&\frac{e2g\hbar}{m^{*}2\pi}
\sum_{n=0}^{n^{N}_{max}}
\int A_{C1}^{N}(\epsilon_{z})B_{N,N+1}(\epsilon_{z})  
T_{N+1}(\epsilon_{z})\nonumber\\ 
&\times&\left[f_{\epsilon_{\omega_{N}}}(\epsilon_{z})
-f_{\epsilon_{F}}(\epsilon_{z}+eV)\right]
d\epsilon_{z}, \label{THM}
\end{eqnarray}
In these expressions 
 $f_{\epsilon_{\omega_{i}}}(\epsilon_{z})=
(1+e^{\frac{(\epsilon_{\omega_{i}}-\epsilon_{z}
-\epsilon(n))}
{k_{B}T}})^{-1}$ are the Fermi functions,
$\epsilon(n)=\hbar\omega_{c}(n+\frac{1}{2})$,
$\omega_{c}=\frac{eB}{m^{*}}$ is the cyclotron
frequency, 
$n$ runs over all Landau levels below the Fermi
energy of the region (emitter
or $i$-th well)  from which the electrons are tunneling. The Fermi
energies, $\epsilon_{\omega_{i}}$,
must be calculated selfconsistently. 
Also, $g=\frac{eBS}{2\pi\hbar}$ is the factor of degeneracy for a given
magnetic field B and area S, 
$j$ labels the resonant state
in each well $i$ with energy $\epsilon_{Cj}^{i}$, $m$ is the total
number of resonant states within the well contributing to the
current and 
$B_{i,i+1} = k_i/(w + \alpha_i^{-1} + \alpha_{i+1}^{-1})$,
where $k_{i}$ and $\alpha_{i}$ are the wave vectors in the
wells and the barriers, respectively. These depend on the local electric 
field. Finally, $T_{i}= 16k_{i-1}k_{i}\alpha_{i}^{2}e^{-2\alpha_{i}d}
(k_{i-1}^{2}+\alpha_{i}^{2})^{-1}(k_{i}^{2}+\alpha_{i}^{2})^{-1}$ is the 
dimensionless transmission probability through the $i$th barrier, and 
$w$ and $d$ are the well and barrier widths respectively. 
\vspace{1cm}
\begin{figure}[h]
\centerline{\epsfxsize=0.45\textwidth
\epsfbox{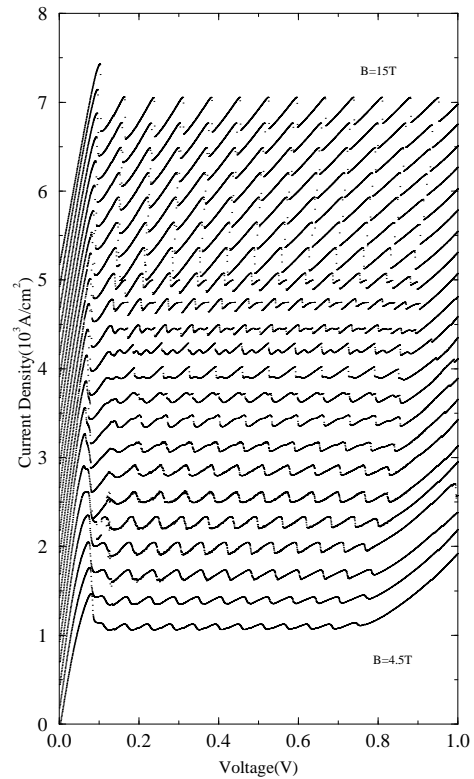}
}
\caption{I--V curves for different magnetic fields in increments of
0.5T. The curves are shifted in intervals of $0.2\times
10^{3}\frac{A}{cm^{2}}$ for clarity.}
\label{fig:schematic1}
\end{figure}
As we mentioned above, we use Lorentzians for the spectral
densities in the wells $A_{Cj}^{i}(\epsilon) = \gamma/[(\epsilon -
\epsilon_{Cj}^{i})^2 +\gamma^2]$.
Although this
assumption can be improved by performing a calculation of the
self-energies including microscopically the scattering
(i.e., including the energy and magnetic field dependence
of the selfenergy), previous
microscopic calculations of impurity scattering in the presence of
magnetic fields have shown this Lorentzian form to be a good
approximation \cite{Fer}. 
On the other hand, our hamiltonian
restricts inter Landau level scattering just within the wells.
The effect of this contribution on the current 
has been shown to be the most
 important one of the possible impurity scattering\cite{Fer}
and interface roughness \cite{Leo}
contributions in a double barrier.\\
Thus, our phenomenological model for
 the scattering accounts for the
main physics: inter-Landau-level scattering at the domain wall (see
below),
and makes the non linear problem tractable numerically.
Also it is important to 
emphasize that the Fermi energies in each well are
variables; i.e, after each tunneling event the lateral energy
$\epsilon(n)$ {\em is not necessarily conserved}.
After reaching a stationary state, at a certain bias,
some amount of charge is accumulated in the well
where the domain wall is formed. Depending on the
degeneracy, it could imply an increase in the maximum Landau level
occupied in one well with respect to its neighboring.
The interwell tunneling then could involve processes where
$\Delta n=n^{i+1}_{max}-n^{i}_{max}\neq$0.\\
We have analyzed a superlattice consisting of 15 wells of GaAs
of 90 $\AA $ thickness and GaAlAs barriers 50 $\AA$ wide.
The emitter doping is $ N_{D}= 2\times 10^{18} cm^{-3}$
and the doping in the wells is $ N_{D}^{w}= 1.5\times 10^{11} cm^{-2} $, 
with $\gamma=4meV$, and $T=0K$.
For zero temperature the expression for the charge density in the $i$-th
well is analytic:
$n_i=\frac{em^{*}\omega_{c}}{\pi^{2}\hbar}\sum_{n=0}^{n^{i}_{max}}
\left[arctg(\frac{\epsilon_{\omega_{i}}-\epsilon(n)\,
-\epsilon_{C1}^{i}}{\gamma})-arctg(\frac{-\epsilon_{C1}^{i}}{\gamma})
\right]$.\\
In Fig. 1 the I--V curve is plotted for different magnetic fields
(the curves are shifted
in the abcissa axis for clarity).
For low magnetic fields, we observe a main peak at low bias coming 
from ground to ground state interwell tunneling, and a sawtooth structure
at high bias due to electric field domains formation. 
As the magnetic
field increases, additional unstable negative differential conductance and
stable branches show up in the current
density. 
\begin{figure}[h]
\centerline{\epsfxsize=0.45\textwidth
\epsfbox{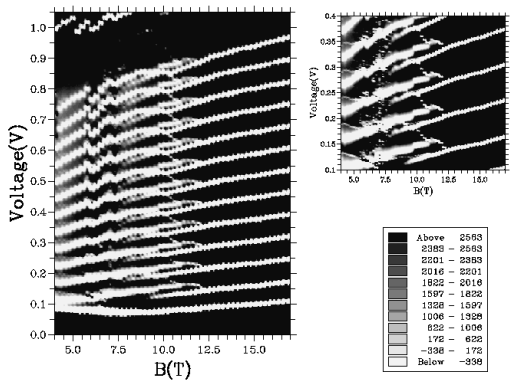}
}
\caption{Differential conductivity versus bias voltage and magnetic field.}
\label{fig:schematic2}
\end{figure}
This extra structure dissappears as the magnetic field is
increased above $\sim 13T$. In order to study  
the range where this
effect manifests itself in more detail
, we plot in Fig. 2 a contour plot of the differential
conductivity as a function of B and the external bias V (negative
differential conductance regions are the white lines).
\vspace{1cm}
\begin{figure}[h]
\centerline{\epsfxsize=0.45\textwidth
\epsfbox{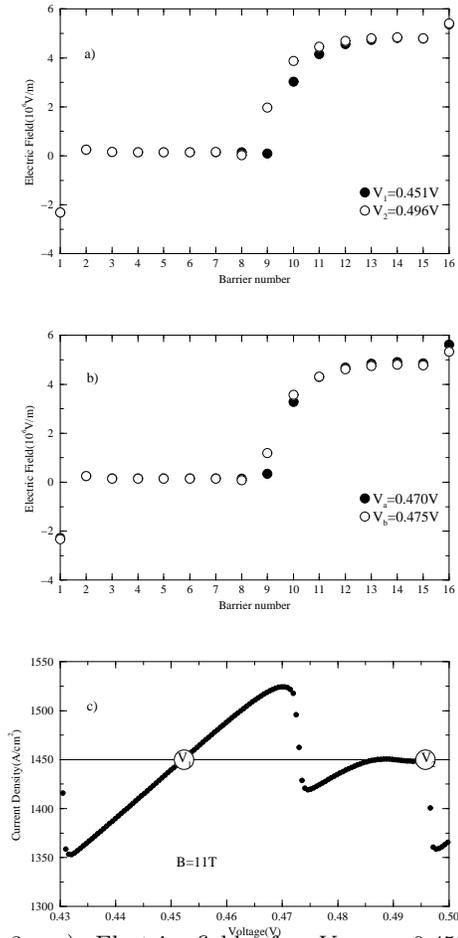}
}
\caption{a) Electric fields for $V_{1}=0.451$Volt and $
V_{2}=0.496$Volt, b)
Electric fields for $V_{a}=0.470$Volt and $ V_{b}=0.475$Volt, c)
Detail of the
 I--V curve for B=11T showing the extra branch.}
\label{fig:schematic3}
\end{figure}
One can see that the additional structure in the current takes
place between 8 and 12.5 T where the differential conductivity shows two 
branches
at fixed magnetic field that repeat periodically. This result is in 
good qualitative
agreement with experiment, that shows this effect up to 19
T \cite{Smi}. At lower magnetic fields 
around 6.5 Tesla, and voltages close to 0.4 Volt,
there is a small structure reflecting the formation of a multistable
solution (see Fig. 2). Our calculations 
have been performed by direct numerical integration of the dynamical
system \cite{Agu} in order to permit a direct comparison with the
experiment 
\cite{Smi}. The experimental I--V
curves have been obtained by sweeping the DC bias only in one
direction --i.e, by 
increasing the DC bias in between the emitter and the
collector. Nonetheless, it is possible to use more complicated numerical
techniques such as the 
numerical continuation method \cite{Agu} in order to
obtain all of the unstable and multistable regions.
The multistable regions for the
current could be observed in the experimental
curves along up and down voltage sweeps.\\ 
We shall concentrate in one region of  bias 
for the current density and fix B=11T in order to
obtain an explanation for these new structures appearing at
intermediate magnetic fields.
We plot in Fig. 3c, a blow up of the current 
for B=11T in the region of voltages where
one third of the wells already belongs to the high field domain.
We have analyzed the electric field distribution
for two different biases that give
the same current density: $V_{1}=0.451$Volt 
and $ V_{2}=0.496$Volt (see Fig. 3a). The first voltage $ V_{1} $,
 corresponds
to a current solution that belongs to the first branch and the second
one $ V_{2}$, corresponds to a solution in the second branch
(Fig. 3c).
\begin{figure}[h]
\centerline{\epsfxsize=0.45\textwidth
\epsfbox{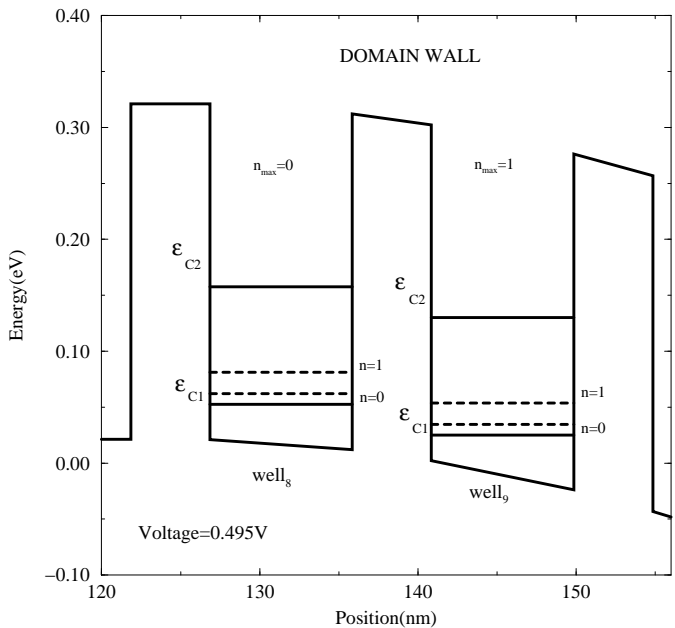}
}
\caption{Detail of the calculated electrostatic profile for 0.495
Volt and B=11T. Dotted lines are the Landau levels. Also the
calculated maximum occupied Landau levels for each well are shown.}
\label{fig:schematic4}
\end{figure}
As we can see, the current density presents a discontinuity
at $ V_{a}=0.47$Volt, and it drops abruptly. Increasing the voltage,
an additional branch shows up.
The current decreases from one branch to 
the next due to the electric field distribution (see in Fig. 3b
the electric field distribution in the neighborhood of the current
discontinuity).
For voltages higher than $V_{a}=0.47$Volt the resonant states in 
neighbor wells at the domain wall
become disaligned in energy, their wavefunctions overlap decreases, and
the current decreases as well. 
The current discontinuity between both branches is due to
 the strongly non linear charge and field distributions.
The electric field
is discontinuous between $V_{a}=0.47$Volt and $V_{b}=0.475$Volt at
the 9th barrier (Fig. 3b).
A detail of the calculated potential profile for a voltage slightly
less than $V_{2}$ (Fig. 4) shows the electrostatic profile at the
domain wall region. This voltage corresponds to the 
formation of the extra branch.
The position of the resonant states --which have been drawn as
discrete ones for simplicity-- and Landau levels energies 
at the domain wall are depicted as continuos and dotted lines
respectively. Also, the calculated maximum Landau level 
 partially occupied in each well, 
$n^{8}_{max}$=0 and $n^{9}_{max}$=1 respectively are indicated. 
These are obtained from the calculated Fermi levels for this voltage.
We conclude that the extra branch of the current is a result
of
 inter Landau level transition at the domain wall 
, $\Delta n=n^{9}_{max}-n^{8}_{max}=1$, between the 8th and 9th wells.
In this situation, the
electrons tunnel from a region with low charge density accumulation
towards the domain wall while changing their maximum
parallel energy to $\hbar\omega_{c}(n^{9}_{max}+\frac{1}{2})$.
As the DC voltage increases, the domain wall moves from one well
to the previous one --in this particular case from the 9th well to the
8th well-- and the structure discussed above repeats periodically
(see Fig.1).
For magnetic fields below 8 Tesla, the motion of the domain wall 
does not involve a change in the maximum
occupied Landau level and there is no inter Landau level
scattering involved. 
In the regime of high magnetic fields, where
the domain boundary enters the magnetic
quantum limit, the lowest Landau level is the
only one occupied in the wells because high degeneracy,
 and we obtain only one branch which repeats
periodically as the domain wall moves.\\
The reason why we do not obtain exactly the same range of magnetic fields as
in  experiment\cite{Smi},
is mainly due to the simplicity of
our model for scattering as we have discussed above. 
We have observed that by increasing the doping in the wells,
 the range of magnetic
fields becomes closer to the experimental one. 
Furthermore, we have not included
 spin-splitting and exchange effects in the calculation.
At high fields, the exchange could be important and the
spin splitting must be considered. This effect
would  also give new features in the non linear current
at high fields. This will be addressed in future work.\\
In summary, we have proposed and solved a dynamical selfconsistent
model for studying the sequential transport through a doped
superlattice in the presence of magnetic fields.
We study the appearance of charge
 instabilities leading to the formation
of electric field domains.
For intermediate magnetic fields, the appearance of
new unstable negative differential conductance regions and new
stable branches in the sawtooth-like I-V curves is discussed.
 These new features in the current can be
explained in terms of inter Landau level scattering occurring at the
domain walls.\\
 One of us (G.P.) acknowledges Prof. K. Von Klitzing 
for addressing us to this problem and for very interesting comments and 
a critical reading of the manuscript. Also  G.P. acknowledges
him and his departament
for their hospitality at MPI in Stuttgart where this work was iniciated.
We acknowledge as well Dr T.
Schmidt for several discussions on his experimental results prior
to publication, Dr. L. Brey for interesting discussions,
Dr. J.P. Rodriguez for a critical reading of the manuscript and Prof. L.L.
Bonilla and Dr. M. Moscoso for collaboration in related topics.\\
(a) Present Address: Department of Applied Physics, Delft University of
Technology, Delft, The Netherlands.

\end{document}